# SOME INNOVATIONS IN DESIGN OF LOW COST VARIABLE COMPRESSION RATIO TWO STROKE PETROL ENGINE.

A.Srinivas[1], G.Venkatasubbaiah[2], P.Venkateswar rao[3] & M. Penchal Reddy[4]

1,2&3.Associate Professor, Department of Mechanical Engineering, Vasavi College of Engineering, Hyderabad-31, India. gvsubbaiah@yahoo.com

4 Asst. Engineer, APGENCO, RTPP, Muddanur, Kadapa dist, India.

*ABSTRACT:*

Historically two stroke engine petrol engines find wide applications in construction of two wheelers worldwide, however due to stringent environmental laws enforced universally; these engines are fading in numbers. In spite of the tight norms, internationally these engines are still used in agriculture, gensets etc. Several designs of variable compression ratio two stroke engines are commercially available for analysis purpose. In this present investigation a novel method of changing the compression ratio is proposed, applied, studied and analyzed. The clearance volume of the engine is altered by introducing a metal plug into the combustion chamber. This modification permitted to have four different values of clearance value keeping in view of the studies required the work is brought out as two sections. The first part deals with the design, modification, engine fabrication and testing at different compression ratios for the study of performance of the engine. The second part deals with the combustion in engine using FLUENT and analysis of exhaust gases. Increase in compression ratio improves fuel efficiency and power output. The novelty in this work is to permit the two wheeler driver to change the compression ratio.

## Introduction:

The two stroke engine is simple to construct. The engine requires two piston strokes or one complete revolution for each cycle. Exhaust ports in the cylinder walls are uncovered by the piston, permitting the escape of exhaust of gases and reducing the pressure in the cylinder. The charge now flows into the cylinder and is compressed in a separate crank case compartment to a pressure greater than atmosphere pressure. Intake ports are uncovered by the piston and the compressed charge flows into the cylinder, expelling most of the exhaust products, however some charge do escape with the exhaust [1, 2]. Comparing four stroke engines, two stroke engines can delivers 50 to 80 % greater power per one piston displacement, high power to weight ratio(twice as many power impulses per cylinder revolutions) piston pin & crank pin experience force in one direction and finally, low cost due to valueless design [3]. The above mentioned features are exploited by the two wheelers stunt drivers, mountain climbing bikes, drag race motor cycles etc.

Worldwide pressure to reduce automotive fuel consumption and exhaust emissions is leading to the introduction of various new technologies for the petrol engine as it fights for market share with the diesel. Variable compression ratio is the technology to adjust internal combustion engine cylinder compression ratio. In a VCR engine high compression ratio is employed for greater efficiency and low load operation, and low compression ratio is employed at full load allowing the turbocharger to work without problem of detonation. So far, variable compression ratio (VCR) engines have not reached the market, despite patents and experiments dating back over decades. VCR technology could provide the key to enable exceptional efficiency at light loads without loss of full load performance [4]. A study on the efficiency and exhaust gas analysis of variable compression ratio spark ignition engine fuelled with alternative fuels reveals that the brake thermal efficiency and volumetric efficiency improved with higher compression ratio [5]. Traditionally, every mechanical element in the power conversion system has been considered as a means to achieve variable compression.

Different methods to obtain variable compression ratio are moving the cylinder head, variation of combustion chamber volume, variation of piston deck height, moving the crankpin or crankshaft, modification of connecting rod geometry [6-11].

1. Moving Head:
By combining head and liners into a semi monoblock construction which pivots with respect to the remainder of the engine, this enables tilting motion to adjust the effective height of the piston crown at TDC. The rotary eccentric can be used to alter the relative position of the two halves of the engine has to overcome the combined inertia of head, liners, supercharger, intercooler and manifolds.

2. Variation of combustion chamber volume:
The volume of the combustion chamber is changed to obtain the variable compression ratio by moving a small secondary piston which communicates with the chamber. The compression ratio is adjusted by using a secondary piston or valve. The device is presented primarily as a means of controlling knock as its dormant state is the high CR condition. It is suggested that the piston could be maintained at an intermediate position, corresponding to the optimum CR for a particular condition, however this would require a finite length bore in which the piston could travel which raises further questions of sealing, packaging and durability.
In another design secondary piston moves continuously at half crankshaft speed and could, potentially, share drive with a camshaft. Phase variation between the secondary pistons and the crankshaft assembly enables the required variation in CR.
Introduction of additional elements may affect the ideal geometry and layout of the valves and ports, hydrocarbon emissions increases due to additional crevice volumes.

3. Variable height piston:
Variation in compression height of the piston offers potentially the most attractive route to the production of VCR engine since it requires relatively minor changes to the basic engine architecture when compared to other options. Unfortunately, it requires a significant increase in reciprocating mass and, more importantly, a means to activate the height variation within a high speed reciprocating assembly. This is typically proposed by means of hydraulics using the engine lubricating oil, however reliable control of the necessary oil flow becomes a major problem.

4. Movement of the crankshaft or crankpins:
Several systems have been proposed which either carry the crankshaft main bearings in an eccentric assembly or move the crankpins eccentrically to affect a stroke change at TDC.

5. Modification of connecting rod geometry:
A popular approach has been to replace the connecting rod with a two piece design in which an upper member connects with the piston while a lower member connects with the crankshaft. By constraining the freedom of the point at which the two members join, the effective height of the connecting rod can be controlled and, hence, the compression volume.

All the above mentioned methods require additional mechanism to vary the compression ratio and make the engine operation complicated. They also results in increased reciprocating masses and cost of the engine. Hence in the present work a simple and cheaper method without using additional mechanism is proposed.

## MATERIALS &METHODS

In the present work an air-cooled, two stroke engine of Bajaj India is modified to operate at different compression ratios. The specifications of the engine are given in the table 1.

Table 1: Test engine specifications:

| Engine type | 2-stroke air cooled single cylinder |
|---|---|
| Engine Power | 3.5 hp (2.576kW)@1500 rpm |
| Scavenging system | Crankcase scavenging |
| Displacement | 145.5 cm$^3$ |
| Compression ratios after modification | 6.4 |
| Bore and stroke | 57 x 57 mm |
| Ignition timing | $22^0$ bTDC |
| Scavenging timing | $122^0$ TDC |
| Combustion chamber | Two-staged semi spherical type |

This engine is a proven one with good track record. A hole is drilled and tapped at a distance of 10mm from the spark plug location and the metal plug (made by EN-8 material) is introduced. The schematic view of the plug along with collars is shown in Figure: 1.

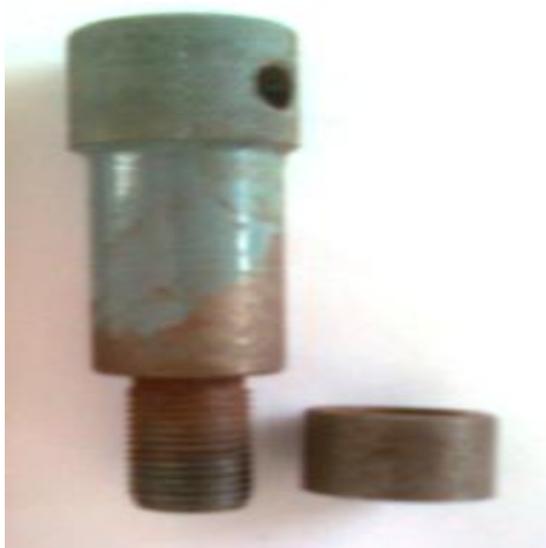

Fig.1: Metal plug and collar

Four variations in the compression ratio are obtainable within the proposed work. The original compression ratio given by the manufacturer cannot be altered at the piston end. However by machining the piston crown it is possible to increase the clearance volume. This machining is restricted by the piston design limitation. The investigation in this line of study has to consider the factor of safety before going for machining. Four collars are made to get four compression ratios apart from existing compression ratio. The compression ratios obtained are 6.4, 6.7, 6.9 & 7.2. Separate levers are provided for shifting gear and operating the clutch and a separate rig is used to mount the engine.

A drum is mounted on the output shaft and is loaded with a brake dynamometer as shown in the test rig in Figure: 2.

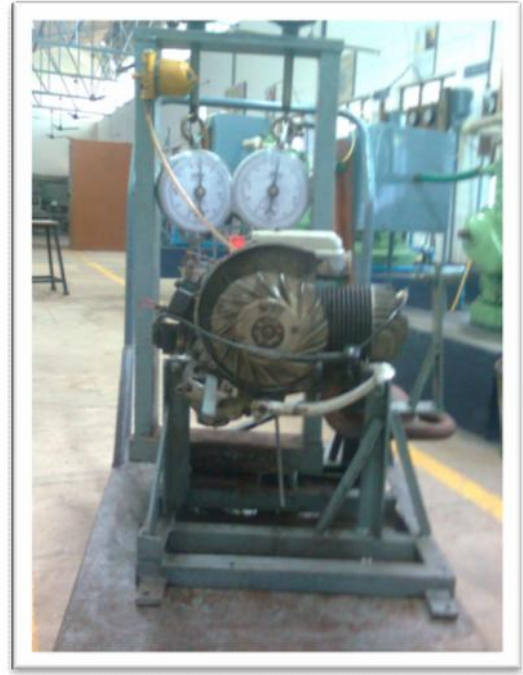

Fig.2: Bajaj scooter engine test rig

The cost of entire test rig is only $150 and this kind of setup can be used for demonstration purposes and for carrying routine experiments as variable compression ratio engine. No attempt is made to patent the design and the future investigations are interested in this area are most welcome to use or modify the proposed retrofit of the engine. The cylinder head with plug in position is shown in Fig. 3, and the inside view of cylinder head with plug in position is shown in Fig. 4. Some clearance volume is occupied by the projection of metal plug in the combustion chamber.

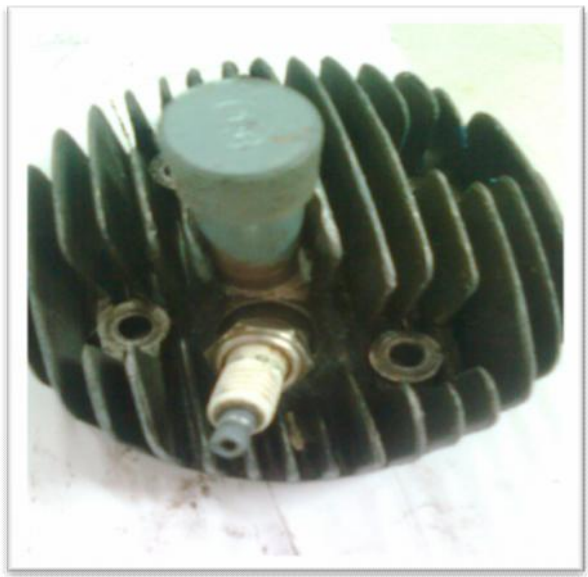

Fig.3: Metal plug & spark plug in position

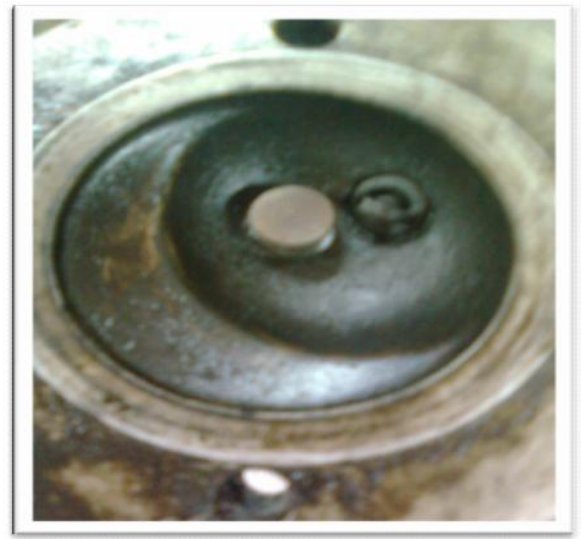

Fig.4: plug and spark plug inside view

The engine is coupled to an eddy current dynamometer to measure brake power. The fuel consumption was measured on mass consumption basis. The clearance volume is found by measuring the length of the pin projected into the combustion chamber
The experiments are carried out at different compression ratios of 6.4, 6.7, 6.9 and 7.2. The load is varied from no load to full load for each compression ratio at constant speed of 1500 rpm of the engine. The variation of total fuel consumption, brake specific fuel consumption, brake thermal efficiency and mechanical efficiency with brake power is studied at constant speed of the engine.

**Results and Discussions:**
The variation of total fuel consumption with brake power at different compression ratios is shown in the fig.5. The total fuel consumption increased with the brake power at all the compression ratios. The total fuel consumption increased up to a compression ratio of 6.9. The improvement in the fuel consumption is considered to be a result of the reduced specific heat ratio of the working gases and increased mechanical loss, cooling loss and time loss.

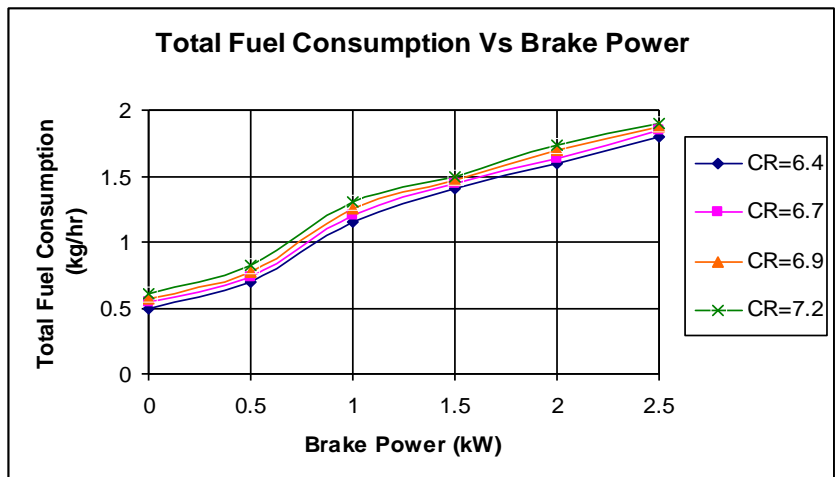

Fig.4 Variation of total fuel consumption with BP at different compression ratios.

The variation of brake specific fuel consumption with brake power at different compression ratios is shown in the fig.6. The BSFC reduced with BP at all compression ratios. The BSFC reduced up to the compression ratio 6.9 but slightly increased with the compression ratio of 7.2.

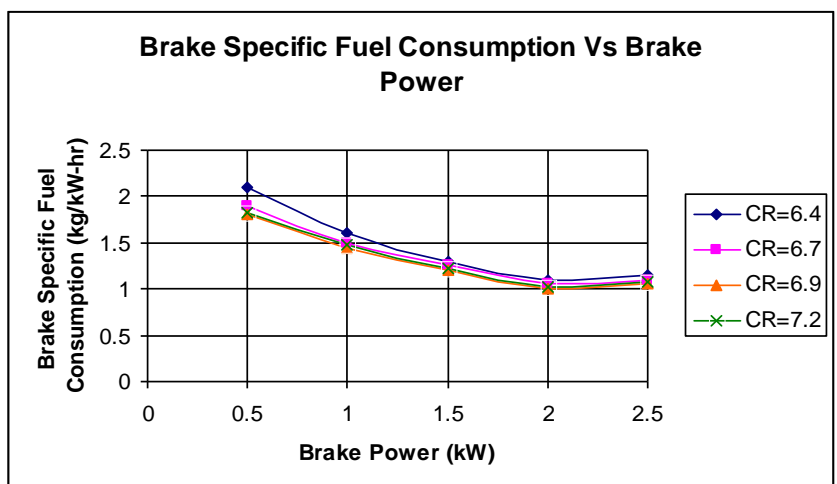

Fig.5 Variation of brake specific fuel consumption with BP

The variation of brake thermal efficiency with brake power at different compression ratios is shown in the fig 6. The results show that the BTE increases with compression ratio. It is due to the working gas temperature affected by an increase in compression ratio. Variation of the working gas specific heat ratio does not prevent the effect of compression ratios.

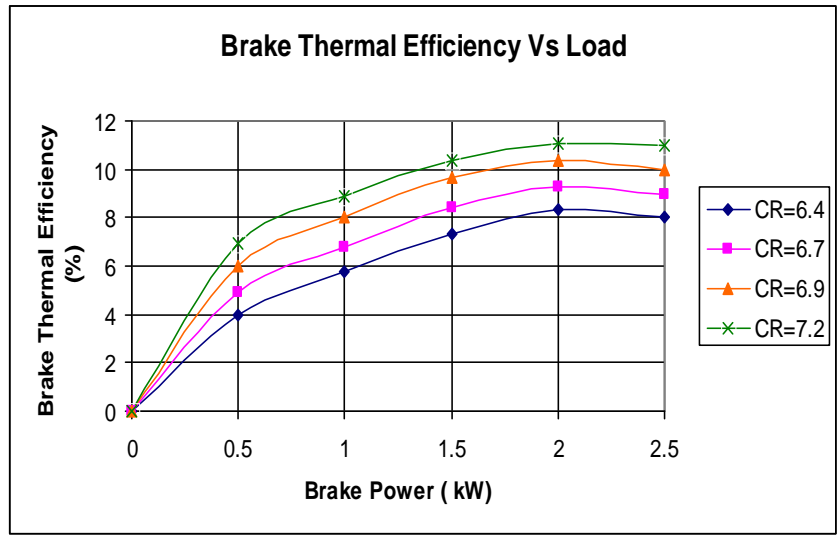

Fig.6: Variation of Brake Thermal Efficiency with compression ratio

The variation of mechanical efficiency with brake power at different compression ratios is shown in the fig 7. The mechanical efficiency increases with compression ratio. The maximum mechanical efficiency is observed with compression ratio of 7.2 at all loading conditions of the engine.

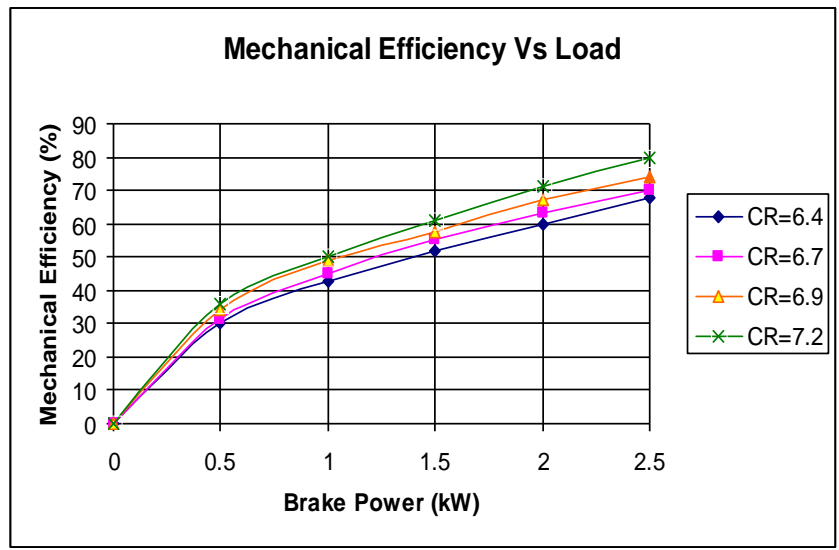

Fig.7: Variation of Mechanical Efficiency with compression ratio

## Conclusions:

The significant conclusions from the present work are summarized as follows.
1. The compression ratio is varied by using a simple.
2. The total fuel consumption increased with the compression ratio.
3. The specific fuel consumption reduced with compression ratio.
4. The brake thermal efficiency increased with compression ratio.
5. The mechanical efficiency increased with compression ratio
6. No elaborate setup is required to change the compression ratio and new design enables the driver to operate at compression ratio of his choice based on terrain he wishes to drive. A knurled head is provided to the pin for easy change.